\documentclass[doublecol]{epl2}
\usepackage{textcomp}
\usepackage{amsmath}
\usepackage{amssymb}
\usepackage{graphicx} 
\usepackage{hyperref}
\usepackage{color}

\newcommand{\ket}[1]{\left|#1\right\rangle}
\newcommand{\bra}[1]{\left\langle #1\right|}

\newcommand{\prm}{\mathrm{p}}

\newcommand{\Trace}{\operatorname{Tr}}

\newcommand{\mean}[1]{\left\langle #1\right\rangle}

\newcommand{\nep}{\textrm{e}}

\newcommand{\Ham}{\widehat{H}}
\newcommand{\Uevol}{\widehat{U}}

\newcommand{\calH}{\mathcal{H}}

\newcommand{\calI}{\mathcal{I}}
\newcommand{\opbfS}{\widehat{\bf S}}
\newcommand{\opS}{\widehat{S}}
\newcommand{\opm}{\widehat{m}}

\newcommand{\Real}{\Re\textrm{e}\,}
\newcommand{\Aimag}{\Im\textrm{m}\,}

\begin{document}

\title{Thermalization in a periodically driven fully-connected quantum Ising ferromagnet}

\author{Angelo Russomanno\inst{1,2,3}\and Rosario Fazio\inst{5,6}\and Giuseppe E. Santoro\inst{2,3,4}}
\shortauthor{A. Russomanno \etal}
\institute{
\inst{1} Department of Physics, Bar-Ilan University (RA), Ramat Gan 52900, Israel\\
\inst{2} SISSA, Via Bonomea 265, I-34136 Trieste, Italy\\
\inst{3} CNR-IOM Democritos National Simulation Center, Via Bonomea 265, I-34136 Trieste, Italy\\
\inst{4} International Centre for Theoretical Physics (ICTP), P.O.Box 586, I-34014 Trieste, Italy\\
\inst{5} Scuola Normale Superiore, Piazza dei Cavalieri 7, I-56127 Pisa, Italy\\
\inst{6} Centre for Quantum Technologies, National University of Singapore, 3 Science Drive 2, 117543 Singapore
}

%
%

\abstract{
By means of a Floquet analysis, we study the quantum dynamics of a fully connected Lipkin-Ising ferromagnet in
a periodically driven transverse field showing that thermalization in the steady state is intimately connected to properties 
of the $N\to \infty$ classical Hamiltonian dynamics. 
When the dynamics is ergodic, the Floquet spectrum obeys a Wigner-Dyson statistics and the system satisfies the 
eigenstate thermalization hypothesis (ETH):
Independently of the initial state, local observables relax to the $T=\infty$ thermal value, and  Floquet states are 
delocalized in the Hilbert space.  On the contrary, if the classical dynamics is regular no thermalization occurs.
We further discuss the relationship between ergodicity and dynamical phase transitions, and the relevance of our 
results to other fully-connected periodically driven 
models (like the Bose-Hubbard), and possibilities of experimental realization 
{in the case of two
coupled BEC}.  
}
%
%


\maketitle
%
%
{\it Introduction.}
Recent experimental advances in ultra-cold atomic systems~\cite{Greiner_Nat02bis,Kinoshita_Nat06,Jaksch_AP05,Trotzky:NP12} 
and femtosecond resolved spectroscopies~\cite{Enciclopedia:book} have made the study of out-of-equilibrium closed many-body quantum 
systems no longer a purely academic question. The key problem in this context are the properties  
of the final/steady state after the system has undergone
a time-dependent perturbation~\cite{Polkovnikov_RMP11, Reimann_PRL08}.
Depending on the nature of the perturbation, particular aspects acquire a prominent role. 
For a gentle (quasi-adiabatic) driving, the distance of the evolved final state from the instantaneous ground state carries 
precious information on the crossing of quantum critical points~\cite{Zurek_PRL05} and on the accuracy of quantum adiabatic computation~\cite{Farhi_SCI01} and quantum annealing protocols~\cite{Kadowaki_PRE98,Santoro_SCI02,Santoro_JPA06}. 
In the opposite case of a sudden quench, the focus is on the (possible) thermal properties of the steady state.
Thermalization is expected in ``classically ergodic'' systems, where the Hamiltonian behaves as a random matrix~\cite{Haake:book}, 
its eigenstates obey the {\em eigenstate thermalization hypothesis} 
(ETH)~\cite{Deutsch_PRA91,Sred_PRE94,Rigol_Nat,Polkovnikov_RMP11},
and relaxation to the micro-canonical ensemble, with vanishing fluctuations in the thermodynamic limit, follows. 
%
Randomness of the eigenstates implies a strong connection between thermalization and delocalization in Hilbert 
space~\cite{Basko_Ann06,Santos_PRE10,Pal_Huse_PRB10,Canovi_PRB11,Marquardt_PRE12,Polkovnikov_RMP11}. 
 
Our goal is understanding the 
properties of the long-time dynamics 
of a many-body quantum system undergoing a periodic driving. 
The interest in periodically driven systems --- a long-standing topic in quantum chaos of small quantum 
systems~\cite{Casati_PTPS89} --- 
has risen again vigorously only quite recently, with a focus on many-body dynamics~\cite{Russomanno_PRL12,
 Bastidas_PRB12,Lazarides_PRL14,Lazarides_PRE14,Abanin_AP15,
 huse_PRE14,Gong_PRL_13,Engelhardt_PRE13,Khripkov_PRE13}
 {and its properties of stability and ergodicity~\cite{Emanuele_2014:preprint}}.
In a recent work~\cite{Russomanno_PRL12} it was proposed that a periodically driven closed quantum system might display,
in the thermodynamic limit, a tendency towards a fully coherent ``periodic steady state'' 
--- a kind of  diagonal ensemble~\cite{Rigol_Nat,Polkovnikov_RMP11} 
 for periodically driven systems ---, 
where destructive interference effects average to zero for long times the transient fluctuations induced by off-diagonal Floquet matrix elements. %
This effect has been explicitly demonstrated on a periodically driven quantum Ising chain, as a 
direct consequence of the smooth continuous nature of the Floquet quasi-energy spectrum.
A quantum Ising chain, being integrable, does not show thermalization: 
The energy per site results from a GGE average~\cite{Jaynes_PR57,Rigol_PRL07,Lazarides_PRL14,Russomanno:phdthesis} and stays
always well below the infinite temperature ($T=\infty$) value~\cite{Russomanno_PRL12}.  
On the contrary, one might conjecture~\cite{Russomanno_PRL12,Russomanno:phdthesis} that when a classically ergodic system is periodically driven, the ``steady state'' would show thermalization to the $T=\infty$ ensemble. 
{This is indeed shown in recent works~\cite{Rigol_PRX14,Abanin_AP15} on two
nonintegrable periodically kicked spin chain models, 
consistently with the Floquet states obeying ETH at $T=\infty$~\cite{huse_PRE14}.}
The same phenomena are observed in Ref.~\cite{Lazarides_PRE14} where the case of interacting hard-core bosons is considered
and the Floquet states are shown to obey ETH being random superpositions of unperturbed eigenstates.
Analogous conclusions for a Bose-Hubbard chain are reported in Ref.~\cite{Dong_14:preprint}.
These results call for a more detailed scrutiny of the relation between integrability and thermalization in 
periodically driven quantum many-body systems.

%
In the present letter we address this relation by studying a periodically driven fully-connected quantum Ising ferromagnet:
a very clear prototype model whose long-time dynamics can be analyzed reliably up to the thermodynamic limit.
The very rich phenomenology we are going to describe should occur also in a driven two-mode Bose-Hubbard model, 
whose Hamiltonian is equivalent to a fully-connected spin system~\cite{Khripkov_PRE13} and is
experimentally feasible by modulating 
the inter-well barrier height in a double-well BEC realization~\cite{Albiez_PRL05}. 
%
Systems of this kind can be mapped, for large $N$, onto large-$S$ quantum spins with effective Planck's constant $\hbar/N$. 
Hence, for $N=\infty$  their dynamics is that of a one-dimensional classical {\em non-linear} Hamiltonian system $\calH(Q,P,t)$ \cite{Sciolla_JSTAT11}. 
%
At variance with the perfectly regular classical dynamics observed after a quantum quench~\cite{Sciolla_JSTAT11, Mazza_PRB12}, 
we will show how rich is the periodic driving case: we can see classically regular motion, chaos and even 
full ergodicity by choosing appropriately the parameters of the driving.
We will solve numerically the Schr\"odinger equation at finite ``large'' $N$.
In both classically ergodic and regular cases, the intensive observables relax to a steady periodic regime: stroboscopic time
fluctuations vanish in the large-$N$ limit.
However, the classically ergodic cases are very different from the regular ones: while the latter show a sensitivity to the initial state 
and never thermalize, the former effectively thermalize towards a $T=\infty$ ensemble. 
This is a consequence, as we will show, of the Floquet states obeying ETH at $T=\infty$ and being delocalized in the Hilbert space.

%
{\it Model.} Here we focus on a fully-connected quantum Ising ferromagnet with a periodically driven 
transverse field, the smooth-driving counterpart of the kicked-top of Ref.~\cite{Haake_ZPB86}. 
Consider $N$ spin-1/2, $\opbfS_{i=1\cdots N}$, and the total spin operator $\opbfS=\sum_i \opbfS_i$. 
The fully connected $\prm$-spin transverse field quantum Ising ferromagnet  \cite{Bapst_JSTAT12} is written as 
$\Ham_{\prm}(t) = - (N J/2) \, \opm_z^{\prm} - N \Gamma(t) \, \opm_x$,
where $J$ is the longitudinal coupling, $\Gamma(t)$ is a (time-dependent) transverse field, 
and  $\opm_{x/z} = 2\opS^{x/z}/N$ are rescaled magnetization operators.
$\Ham_{\prm}(t)$ commutes with $\opbfS^2$, and the equilibrium ground state 
is a state of {\em maximum spin}, $S=S_{max}=N/2$, belonging to the $(N+1)$-dimensional multiplet 
of spin eigenstates $|S=N/2,M\rangle$. 
Since $[\Ham_{\prm}(t),\opbfS^2]=0$, the Schr\"odinger dynamics starting from an initial state 
$|\psi_0\rangle$ with $S=N/2$ will always remain in that sector.
If $m=2M/N$ are the eigenvalues of $\opm_z$, the multiplet of interest has $m=-1 +2j/N$ with $j=0,\cdots, N$; 
we denote it as $|S=N/2,M\rangle_z \to |m\rangle$.
The case $\prm=2$ corresponds to the Ising-anisotropic version \cite{Botet_PRB83} of the so-called 
Lipkin model \cite{Lipkin_NucPhys65} (see also Refs.~\cite{Latorre_PRA05,Latorre_PRL07}):
\begin{equation}  \label{Lipkin-model:eqn}
  \Ham_{\prm=2}(t)= -\frac{2J}{N}\sum_{i,j}^N \opS_{i}^z \opS_j^z - 2 \Gamma(t) \, \sum_{i}^N \opS_i^x \;.
\end{equation}
When $\Gamma$ is constant, $\Ham_{\prm=2}$ has a quantum critical point (QCP) 
at $\Gamma_c/J=1$ separating a large-$\Gamma$ quantum paramagnet from a low-$\Gamma$ ferromagnet; 
for $\prm>2$ the transition is first order \cite{Jorg_EPL10}.
%
%
%
The non-equilibrium quantum dynamics of these models has so-far been discussed
%
in the cases of quantum annealing~\cite{Kadowaki_PRE98,Santoro_SCI02,Santoro_JPA06}   
across the QCP \cite{Das_PRB06,Caneva_PRB08,Jorg_EPL10,Bapst_JSTAT12}, 
%
%
%
%
and of a sudden quench of $\Gamma(t)$ \cite{Sciolla_JSTAT11,Mazza_PRB12}, 
in the context of dynamical phase transitions. 
%
%
Here we will consider its non-equilibrium coherent dynamics  
under a periodic transverse field, 
more specifically $\Gamma(t) = \Gamma_0 + A \sin{(\omega_0 t)}$.
The properties of a fully-connected spin chain undergoing a smooth periodic driving have been discussed
with the rotating wave approximation in the context of non-equilibrium phase transitions in Ref.~\cite{Engelhardt_PRE13} 
and from the perspective of many-body coherent destruction of tunneling in the limit of high driving frequency in
Ref.~\cite{Gong_PRL_13}; here we take a different
point of view and focus on the regularity/ergodicity properties of the quantum many body system.

To discuss its exact quantum dynamics, we have to expand
the state $|\psi(t)\rangle$ on the $(N+1)$-dimensional basis $|m\rangle$ 
as $|\psi(t)\rangle = \sum_{m} \psi_m(t) | m \rangle$, 
the Schr\"odinger equation reads:
\begin{equation} \label{Schroedinger:eqn}
i\hbar \frac{\partial}{\partial t} \psi_m = -\frac{N}{2} J m^{\prm} \psi_m 
- \frac{N}{2} \Gamma(t) \sum_{\alpha = \pm 1} h_{m}^{\alpha} \psi_{m+\alpha \frac{2}{N}} \;, 
\end{equation}
with $h_{m}^{\pm}=\sqrt{1-m^2 + 2(1 \mp m)/N}$. 
%
The system becomes increasingly classical for $N\to \infty$: indeed 
the commutator $\left[ \opm_x, \opm_y \right] = i (2/N) \opm_z$ vanishes in that limit. 
A careful semi-classical analysis \cite{Sciolla_JSTAT11,Bapst_JSTAT12} 
reveals that the expectation values of the magnetization are effectively described, for $N=\infty$, 
by a one-dimensional classical Hamiltonian of the form  
\begin{equation} \label{classicham:eq}
\calH_{\prm}(Q,P,t) = - \frac{J}{2} Q^{\prm} - \Gamma(t) \; \sqrt{1-Q^2} \; \cos{(2P)} \;,
\end{equation}
with the identification $\langle \psi(t) | \opm_z | \psi(t) \rangle \rightarrow Q(t)$, 
and $\langle \psi(t) | \opm_x| \psi(t) \rangle \rightarrow \sqrt{1-Q^2(t)} \; \cos(2P(t))$.  
%
After a sudden quench of $\Gamma$, 
energy conservation gives an integrable $\calH_{\prm}(Q,P)$; 
under a periodic $\Gamma(t)$, on the contrary, classical chaos in the $(Q,P)$ phase 
space can emerge ~\cite{Berry_regirr78:proceeding}. 
In the rest of the paper we will focus on the case $p=2$, whose theoretical and experimental importance
relies also in the fact that Eq.~\eqref{Lipkin-model:eqn} describes the exact dynamics~\cite{Khripkov_PRE13} and
Eq.~\eqref{classicham:eq} with $p=2$ the mean field dynamics~\cite{Smerzi_PRL97} of two coupled-trapped 
Bose-Einstein Condensates under a time-periodic modulation. 

%

%
{\it Floquet analysis.}
The natural framework to study the time evolution of a periodically driven quantum system is the Floquet theory~\cite{Shirley_PR65, Grifoni_PR98}. 
It states that there exists a basis of solutions of the Schr\"odinger equation which are periodic ``up to a phase factor'' 
$\nep^{-i\mu_\alpha t}\ket{\phi_\alpha(t)}$,  where $\ket{\phi_\alpha(t)}=\ket{\phi_\alpha(t+\tau)}$, 
$\tau=2\pi/\omega_0$ being the period. 
The Floquet quasi-energies $\mu_\alpha$ and modes $\ket{\phi_\alpha(t)}$ 
are obtained by diagonalizing the evolution operator over one period
$\Uevol(\tau) \ket{\phi_{\alpha}(0)} =  \nep^{-i\mu_{\alpha}\tau} \ket{\phi_{\alpha}(0)}$,
with $\mu_{\alpha}\in [-\omega_0/2,\omega_0/2]$.
If we consider the stroboscopic dynamics at times $t_n=n\tau$, since $\Uevol(n\tau)= \Uevol^n(\tau)$, 
the Schr\"odinger evolution is completely determined by $\ket{\phi_\alpha(0)}$ and $\mu_{\alpha}$:
the state can be written as 
$\ket{\psi(n\tau)}=\sum_\alpha \nep^{-in\mu_\alpha\tau} R_{\alpha} \ket{\phi_\alpha(0)}$, 
where $R_\alpha=\langle\phi_\alpha(0)\ket{\psi_0}$. 
%
We focus on the $\prm=2$ case, considering intensive observables like the energy-per-site
\begin{equation} \label{energy_per_site:eqn}
  e_{\psi_0}(n\tau) = \frac{1}{N}\bra{\psi(n\tau)}\Ham_{\prm=2}(0)\ket{\psi(n\tau)} \;.
\end{equation}
Provided the Floquet spectrum is non-degenerate (which we have verified numerically), we can easily evaluate 
the finite-$N$ stroboscopic infinite-time average of $e_{\psi_0}$ and the corresponding squared fluctuations 
$\delta{e}_{\psi_0}^2= \overline{e^2_{\psi_0}} -\overline{e}^2_{\psi_0}$: 
{
\begin{eqnarray} \label{aver:eqn}
\overline{e}_{\psi_0} &=&\lim_{n\to\infty}\frac{1}{n}\sum_{k=0}^{n-1} e_{\psi_0}(k\tau)=
\sum_{\alpha} \left|R_\alpha\right|^2 e_{\alpha\alpha} \\
\label{fluc:eqn}
\delta{e}_{\psi_0}^2 
   &=&
\sum_{\alpha\neq\beta} \left|R_\alpha\right|^2 \left|R_\beta\right|^2 |e_{\alpha\beta}|^2 \;,
\end{eqnarray}
}
%
%
where {$e_{\alpha\beta}=\bra{\phi_\alpha(0)}\Ham_{\prm=2}(0)\ket{\phi_\beta(0)}/N$}. 
{We can think of $\overline{e}_{\psi_0}$ as} a ``Floquet diagonal ensemble average'' \cite{Rigol_Nat,Polkovnikov_RMP11,Russomanno:phdthesis}.
%

\begin{figure}
\includegraphics[width=85mm]{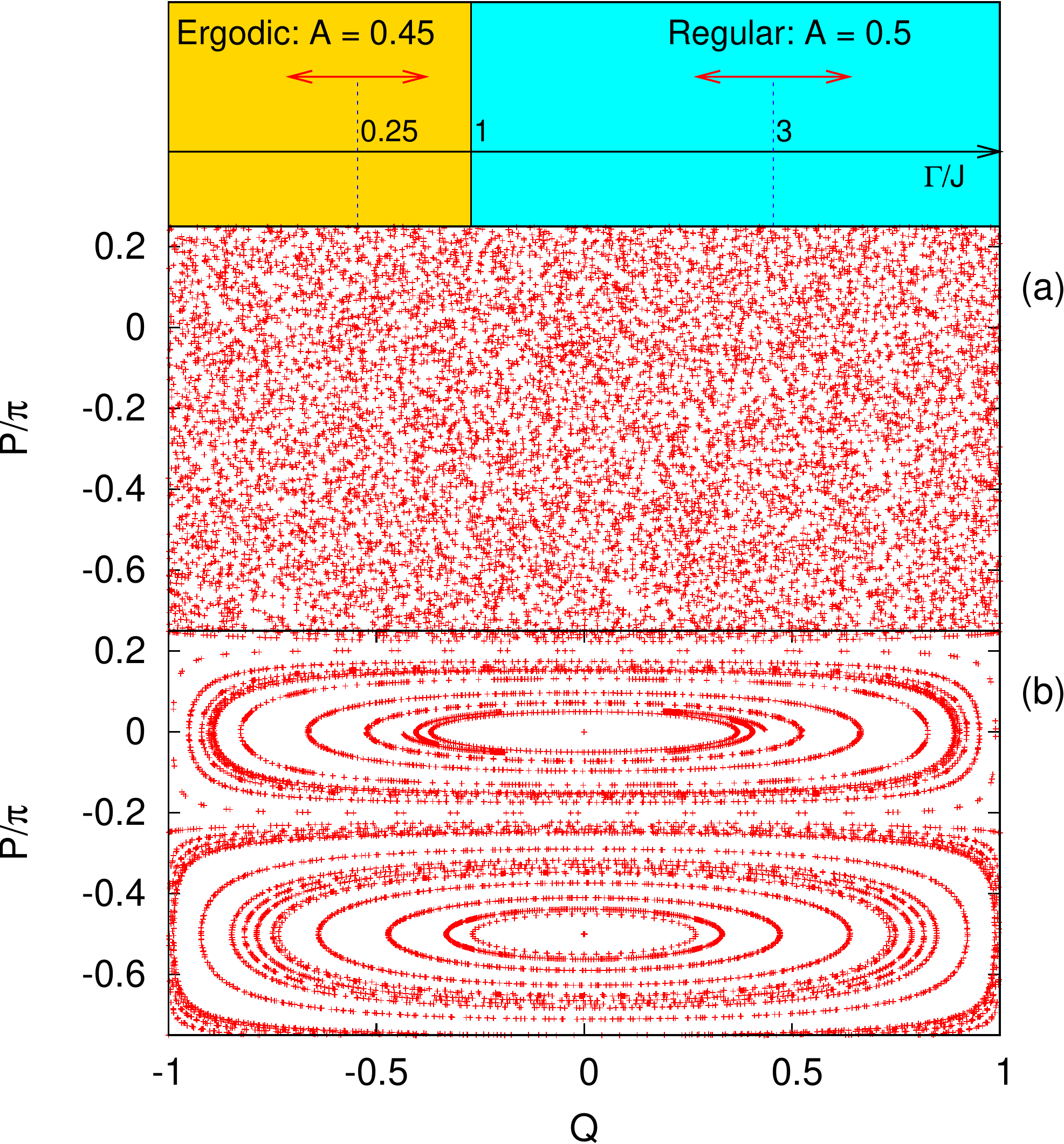}
\caption{(Top) Schematic equilibrium phase diagram with the two representative driving cases illustrated.
(a) and (b) Poincar\'e sections of the classical dynamics at stroboscopic times $t_n=n\tau$ in the $(Q,P)$ phase space.
The parameters are $\Gamma_0/J=0.25$, $A/J=0.45$ for (a),
and $\Gamma_0/J=3$, $A/J=0.5$ for (b), with $\omega_0/J=2$ and $\Gamma(t)=\Gamma_0 + A \sin{(\omega_0 t)}$.
Chaos is fully developed in (a); the motion is regular in (b).}
\label{Poincare-ergodic:fig}
\end{figure}

%
{\it Results.}
The phenomenology of the driven model is quite rich, depending on $\Gamma_0$, $A$ and $\omega_0$. 
It will be helpful to use the equilibrium phase diagram as a guide 
(although the dynamics has no strict relation to the different equilibrium phases).
When $\Gamma_0<\Gamma_c=J$, a classical hyperbolic point at $(Q,P)=(0,0)$~\cite{Sciolla_JSTAT11} 
makes the system prone to chaos 
even with a small $A$~\cite{Berry_regirr78:proceeding}: 
the stroboscopic Poincar\'e sections in Fig.~\ref{Poincare-ergodic:fig}(a) 
show an instance of ergodic phase space with fully developed chaos.
At the quantum level~\cite{Haake:book},  the corresponding distribution of Floquet quasi-energy spacings $P(S)$
--- with $S_\alpha=\rho(\mu_\alpha)(\mu_{\alpha+1}-\mu_\alpha)$, where $\rho(\omega)=\mean{\sum_\alpha\delta(\omega-\mu_\alpha)}_{\Delta}$ 
is the density of quasi-energies smoothed over a ``mesoscopic'' scale $\Delta$~\cite{Haake:book,Berry_LH84} --- 
is well described by an orthogonal Wigner-Dyson distribution~\cite{Haake:book}
$P_{\rm WD}(S)=\frac{\pi}{2}S\exp\left(-\frac{\pi}{4}S^2\right)$.
When $\Gamma_0>\Gamma_c$, inside the equilibrium paramagnetic phase, on the contrary, the classical motion tends 
to be more regular, and a larger $A$ is needed for a substantial chaotic component 
in phase space: for $\Gamma_0=3J$, an $A/J=0.5$ still shows very regular classical motion,
Fig.~\ref{Poincare-ergodic:fig}(b), and a Poisson level statistics $P_P(S)=\nep^{-S}$~\cite{Berry_PRS76}.
{In the last case, the Poisson statistics gives rise, for very large $N$, 
to a Floquet spectrum with a non-extensive number of quasi-degeneracies;
nevertheless relaxation to the Floquet-diagonal ensemble and Eqs.~\eqref{aver:eqn} are still valid,
as shown in Refs.~\cite{Reimann_PRL08,Reimann_NJP12} for the analogous case of a quantum quench with
degeneracies in the energy spectrum.}
Results for the regular and ergodic cases
are very similar to those found in the kicked-top problem~\cite{Haake_ZPB86}. 
Here we will focus on these two paradigmatic cases; see Ref.~\cite{Russomanno:phdthesis}
for details about the intermediate situations.
%
%

\begin{figure}
\begin{center}
\includegraphics[width=8.7cm]{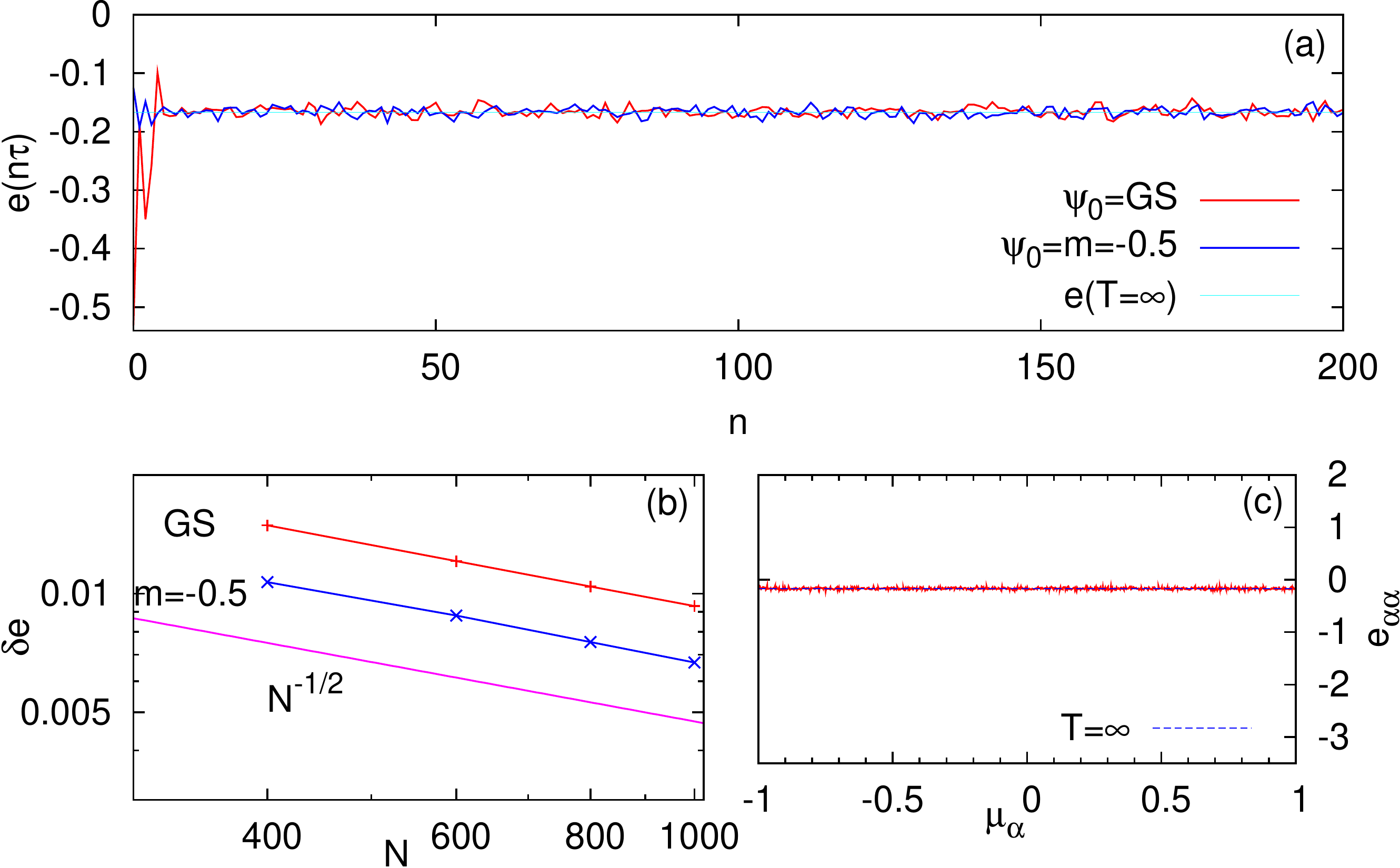}
\end{center}
\caption{(a) Stroboscopic evolution of energy $e_{\psi_0}(n\tau)$ in the ergodic case  $\Gamma_0/J=0.25$, 
$A/J=0.45$, $\omega_0/J=2$, for two different initial states $\psi_0$, 
the symmetry-broken ground state (GS) and an eigenstate $\ket{m}$ of $2\opS_z/N$, {simulations are for $N=800$}.
Here we can see that $e_{\psi_0}(n\tau)$ converges to $e_{T=\infty}$, in panel b) that the 
fluctuations $\delta e_{\psi_0}$ (panel b) decay as $N^{-1/2}$.
(c) $e_{\alpha\alpha}=H_{\alpha\alpha}/N$ (the diagonal Floquet matrix elements)
vs the Floquet quasi-energy $\mu_\alpha$ for $N=800$: $e_{\alpha\alpha}$ is almost constantly equal 
to $e_{T=\infty}$ value, in agreement with $T=\infty$-ETH. 
}
\label{ergo:fig}
\end{figure}

We now turn to the evolution of the observables. 
If $e_{\psi_0}(n\tau)$ relaxes to a periodic steady regime, its asymptotic stroboscopic value must equal
Eq.~\eqref{aver:eqn}, and time fluctuations, Eq.~\eqref{fluc:eqn}, have to vanish. 
In Fig.~\ref{ergo:fig}(a) we show $e_{\psi_0}(n\tau)$ in the ergodic case: 
it clearly relaxes to $\overline{e}_{\psi_0}$. 
Two important points are in order: 
\begin{itemize}
\item[{\em i)}] $\delta e_{\psi_0}$ vanishes as $\sim N^{-1/2}$ (Fig.~\ref{ergo:fig}(b)),
showing indeed relaxation in the thermodynamic limit; 
\item[{\em ii)}] $\overline{e}_{\psi_0}$ is independent of the initial state $\ket{\psi_0}$, 
up to differences of order $N^{-1/2}$, and equal to the $T=\infty$ thermal average 
$e_{T=\infty}=\frac{1}{N(N+1)} \Trace_{S_{\rm max}}\left[\hat{H}(0)\right]$ (the trace is restricted to the $S_{\it max}=N/2$ subspace).
\end{itemize}
%
{This is true for every $\ket{\psi_0}$:
as Fig.~\ref{ergo:fig}(c) shows, the Floquet diagonal terms $e_{\alpha\alpha}$
entering in Eq.~\eqref{aver:eqn}) are all equal to $e_{T=\infty}$, up to fluctuations of order $N^{-1/2}$,
and one can see almost by inspection that property {\em ii)} follows
whatever is the initial state, thanks to the normalization $\sum_\alpha|R_\alpha|^2=1$.
Indeed, when the dynamics
is ergodic, all the Floquet states are equivalent: they are superpositions of energy eigenstates
with random phases and each one is equivalent to the $T=\infty$ thermal ensemble: they behave as
eigenstates of a random matrix~\cite{Berry_LH84,Sred_PRE94,Haake:book,Bohigas_PRL84,Peres_PRA84,Deutsch_PRA91} 
and obey the Eigenstate Thermalization Hypothesis at $T=\infty$.}
Concerning fluctuations,  
we have verified numerically (see Ref.~\cite{Russomanno:phdthesis} and the Supplementary Material)
that the Floquet off-diagonal terms $|e_{\alpha\beta}|$ in Eq.~\eqref{fluc:eqn} scale like $N^{-1/2}$ and, consequently,
so does $\delta e_{\psi_0}$, whatever is the initial state ({property {\em i)}). There is also an analytical argument leading to this scaling.
It relies on the fact that the Floquet states obey $T=\infty$-ETH and are indeed uniform superpositions with random phases
of the eigenstates of the Hamiltonian
\begin{equation}
  \ket{\phi_\alpha} = \frac{1}{\sqrt{N+1}}\sum_{n=1}^{N+1}\nep^{-i\theta_n^\alpha}\ket{n}\,,
\end{equation}
where $\theta_n^\alpha$ are independent random variables uniformly distributed in $[0,2\pi]$. 
Using this formula, the fact that $\ket{n}$ are eigenstates of $\hat{H}(0)$ and the central
limit theorem, it is easy to show that the distributions of the real and the imaginary part of
$e_{\alpha\beta}$ have a variance scaling like $\sim 1/N$ and indeed 
$\delta e_{\psi_0}\sim N^{-1/2}$. 
The detailed derivation is reported in the Supplementary Material.
}
{Thermalization to $T=\infty$ by means of ETH} applies whenever the system is classically 
ergodic and is valid for the energy as well as for all the other intensive observables. 
%
%
\begin{figure}
\begin{center}
\includegraphics[width=8.5cm]{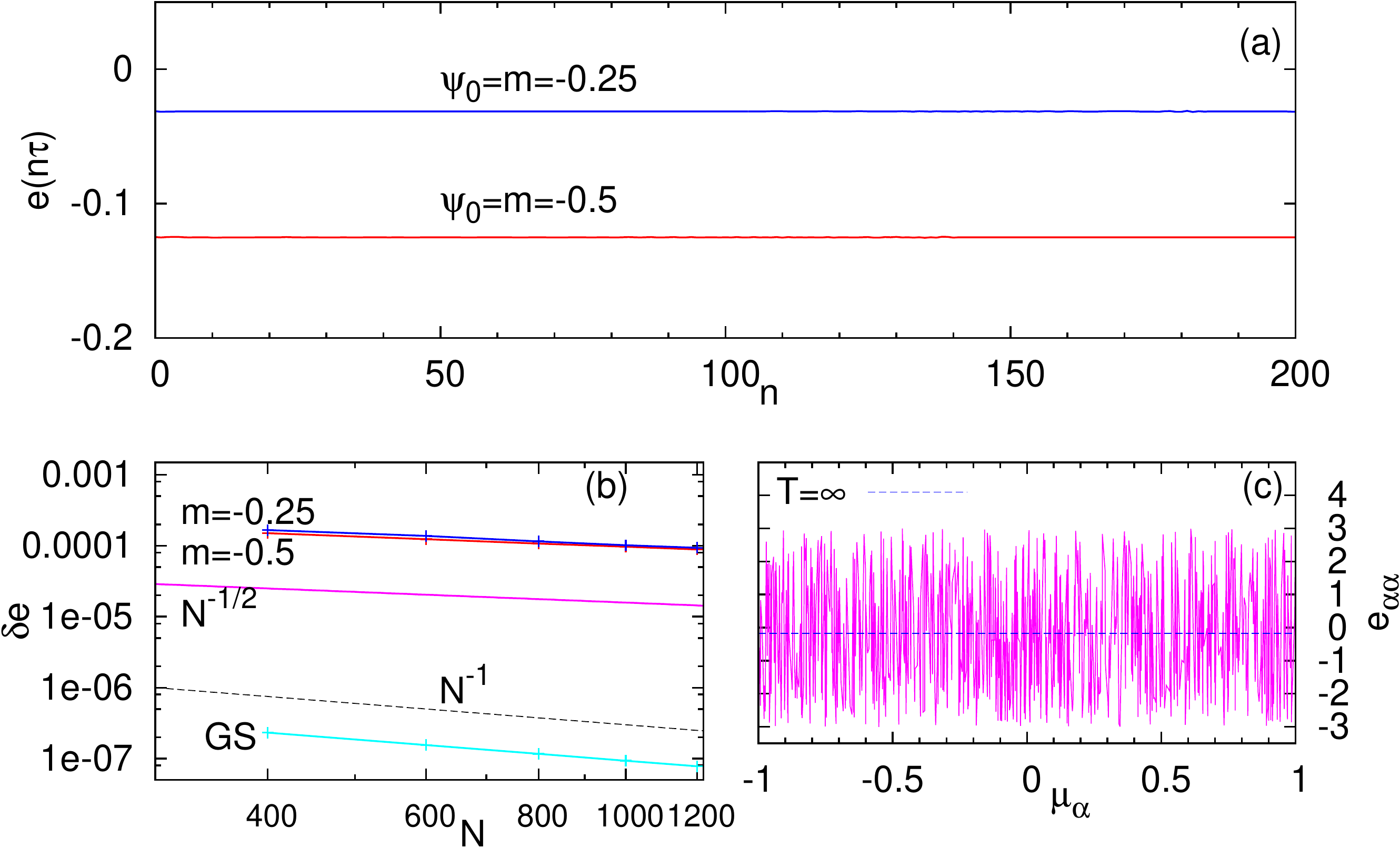}
\end{center}
\caption{
Similar to Fig.~\ref{ergo:fig}, but for the classically regular case $\Gamma_0/J=3$, $A/J=0.5$, $\omega_0/J=2$.
Here the initial states are two different eigenstates $\ket{m}$ of $2\opS_z/N$ and the {ground state}. 
{(In panel a) we omit the results for the GS which show very small fluctuations around $e=-3$).} 
Notice that $\overline{e}_{\psi_0}$ depends on $\psi_0$, {while the (very small) fluctuations decay with $N$ with a
scaling exponent
dependent on $\ket{\psi_0}$ (panel b), and $e_{\alpha\alpha}$ fluctuates wildly with $\mu_{\alpha}$ (panel c).} 
}
\label{regular:fig}
\end{figure}

The physics in the classically regular case is very different:
Fig.~\ref{regular:fig}(a) shows examples of $e_{\psi_0}(n\tau)$:
We see relaxation to the Floquet diagonal ensemble ($\delta e_{\psi_0}$ is practically invisible and scales to $0$),
but the asymptotic value $\overline{e}_{\psi_0}$ strongly depends on $\ket{\psi_0}$. 
The Floquet states behave here very differently: the diagonal terms $e_{\alpha\alpha}$ strongly 
depend on $\alpha$ (see Fig.~\ref{regular:fig}(c)) and there is no ETH. 
{We show numerically in the Supplementary material that also in this case the off-diagonal terms 
$|e_{\alpha\beta}|$ scale on average to zero like $N^{-1/2}$. We also find that these terms show larger fluctuations 
than in the ergodic case; this is the reason why $\delta e_{\psi_0}$ scales, in the regular case, in a less smooth way
with a scaling exponent depending on $\ket{\psi_0}$ (see the bottom-left panel of Fig.~\ref{regular:fig}, where for
the eigenstates $\ket{m}$ of $2\opS_z/N$ we have $\delta e_{m}\sim {N}^{-1/2}$ and for the ground state $\delta e_{\rm GS}\sim N^{-1}$).}
%
\begin{figure}
\includegraphics[width=85mm]{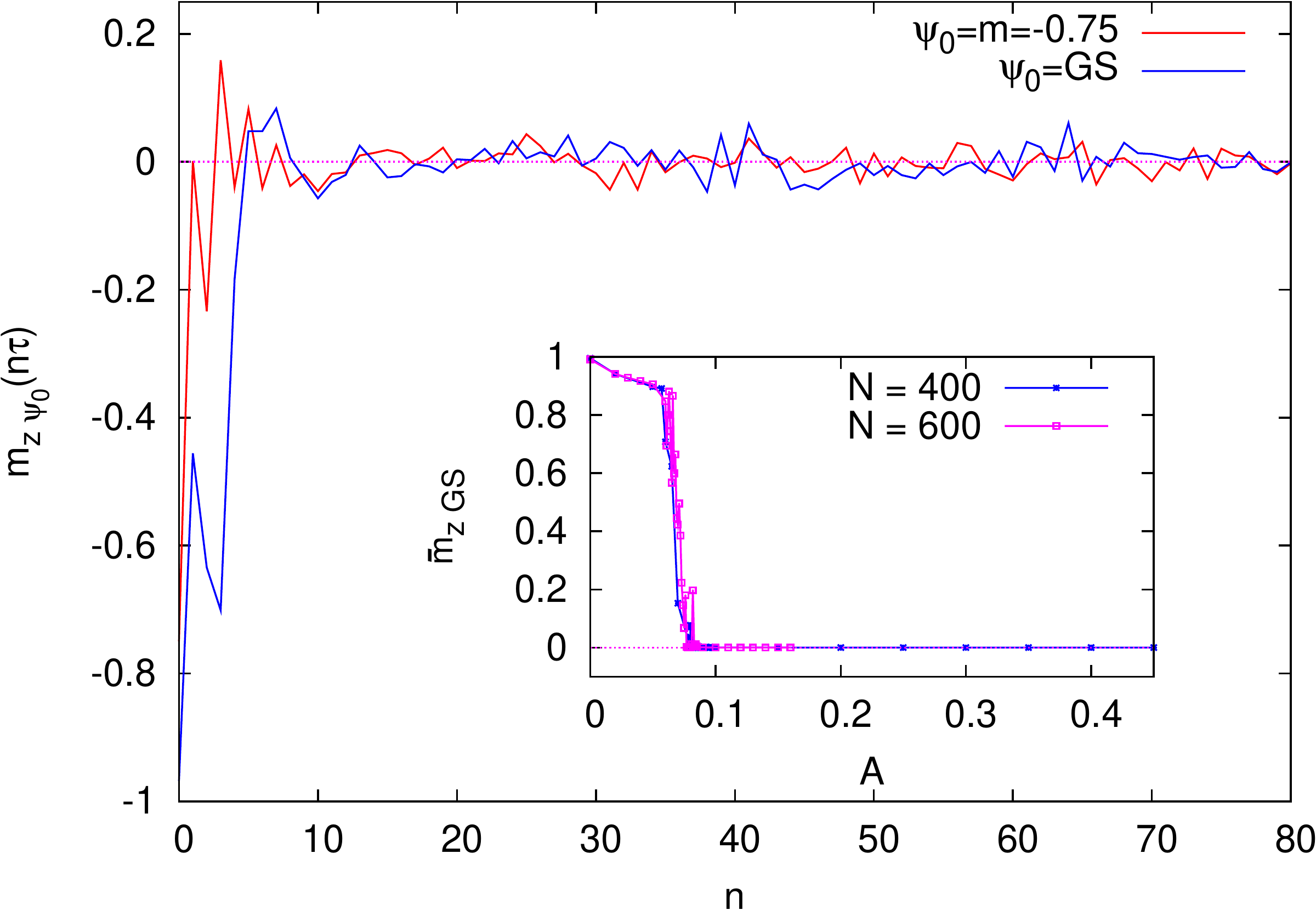}
 \begin{tabular}{cc}
  \hspace{0.4cm}\resizebox{37mm}{!}{\includegraphics{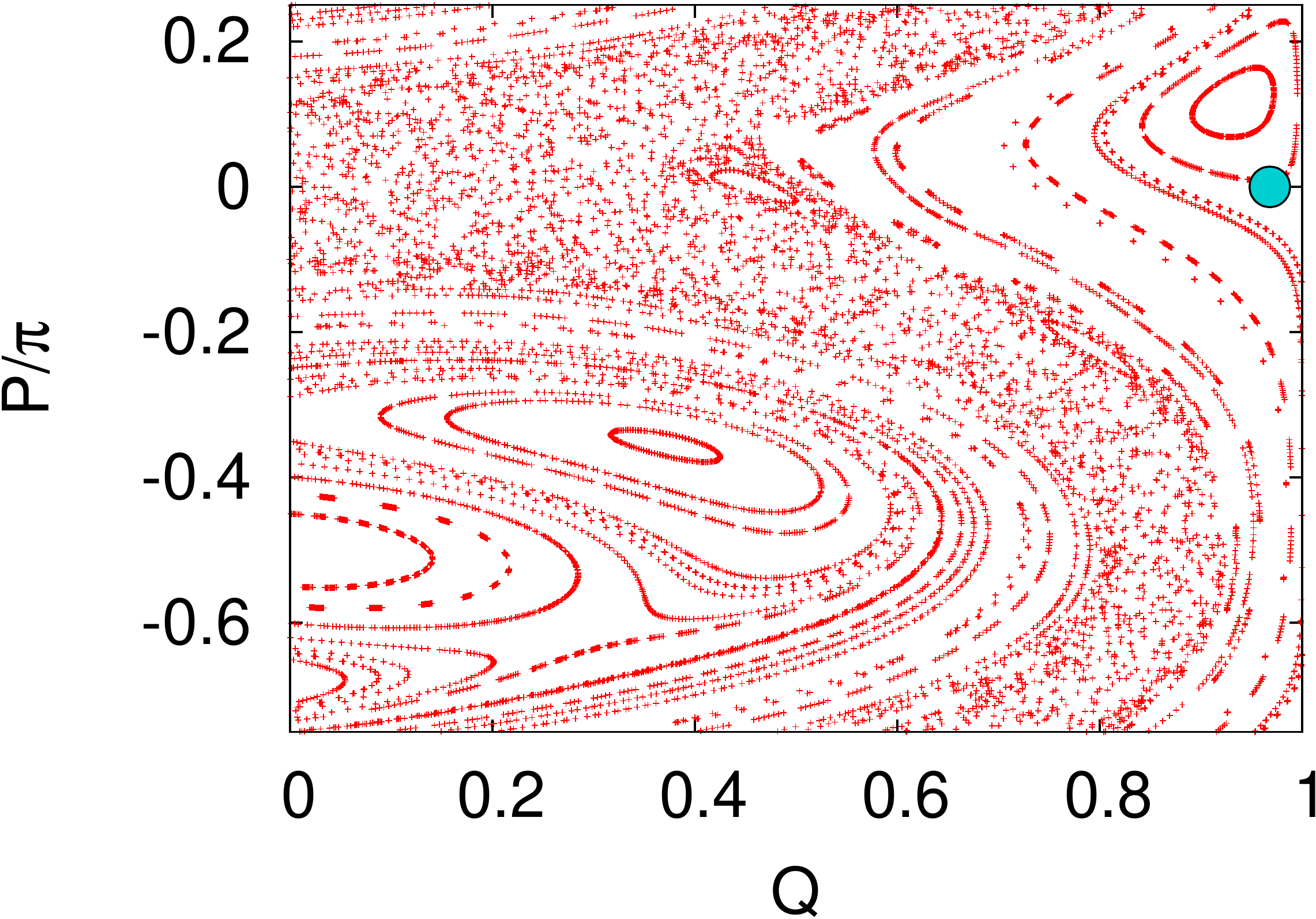}}&
  \hspace{0.cm}\resizebox{37mm}{!}{\includegraphics{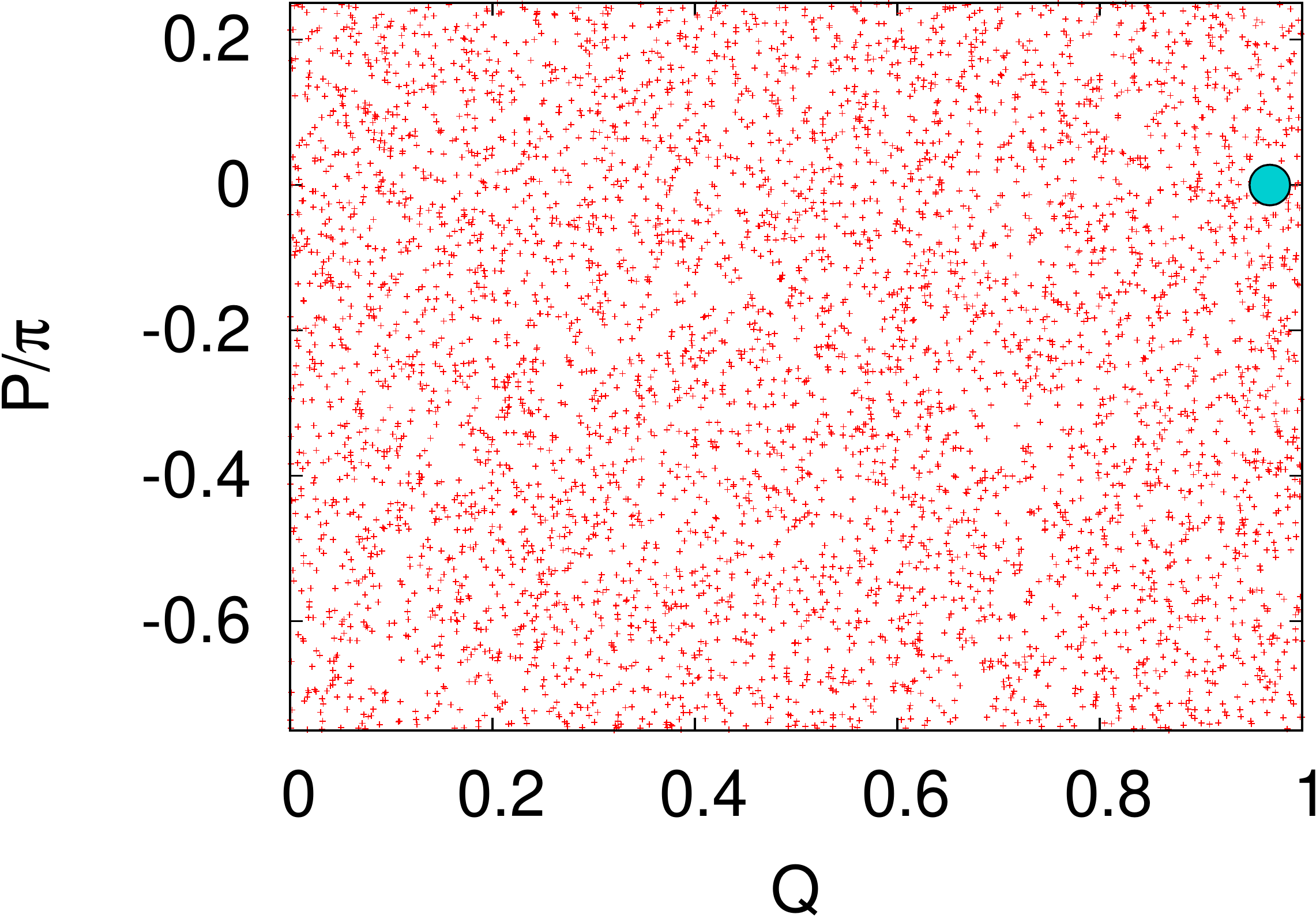}}
 \end{tabular}
\caption{
(Upper panel) Stroboscopic evolution of $m_{z,\psi_0}(n\tau)$ starting from $|\psi_0\rangle=\ket{m=-0.75}$ and $|{\rm GS}\rangle$
in the ergodic case, showing a quick convergence towards $\overline{m}_{z,\psi_0}=0$.
Inset: $\overline{m}_{z,{\rm GS}}$ vs $A$ for $\Gamma_0/J=0.25$ and $\omega_0/J=2$, showing a 
dynamical phase transition {(at $A_c\sim 0.07$)} similar to that found in Ref.~\cite{Sciolla_JSTAT11}.
{(Lower panels) Classical phase space interpretation of the transition: for $A<A_c$ (left panel, $A=0.03$)
the phase space point representative of the ground state is in a regular broken-symmetry region of the phase space; 
for $A>A_c$ (right panel, $A=0.45$) it is in a chaotic region symmetric around $Q=0$.}
}
\label{mz:fig}
\end{figure}

Consistently with this picture, the order parameter stroboscopic time average $\overline{m}_{z\psi_0}$ vanishes 
if ergodicity is at play,  even if $\Gamma(t)$ is always within the equilibrium ferromagnetic phase.
We see this in Fig.~\ref{mz:fig}, where we show ${m}_{z\psi_0}(n\tau)$ for two different
$\ket{\psi_0}$.
Although different in the details, this finding is in line with the ``dynamical transition'' found 
upon quenching from the ferromagnetic phase \cite{Sciolla_JSTAT11,Mazza_PRB12}.
Indeed, by taking $\ket{\psi_0}$ as the broken-symmetry ferromagnetic $|{\rm GS}\rangle$ at $\Gamma_0$
and considering the dependence of $\overline{m}_{z\,{\rm GS}}$ on the driving field amplitude $A$,
we observe a transition 
at a critical value $A_c$ (see the inset) 
{independent of the number of particles $N$. 
As done in Ref.~\cite{Sciolla_JSTAT11} for the case of quantum quench, 
we can give a classical phase space interpretation of this fact. 
In the classical $N\to \infty$ limit the ground state is a point in the phase space; if $\Gamma_0$ is
in the broken symmetry phase this point has coordinates $P=0$ and $Q=\pm\sqrt{1-\Gamma_0^2}$. 
For $A<A_c$ this point falls in regular region of the phase space (bottom left panel of Fig.~\ref{mz:fig}) 
and its subsequent dynamics is trapped in a torus~\cite{Berry_regirr78:proceeding} which is not symmetric 
around $(m_z)_{N\to\infty}=Q=0$. 
Instead, when $A>A_c$, the ground state phase space point
falls in a chaotic region symmetric around $Q=0$ (bottom right panel)
which is ergodically explored by the subsequent dynamics, and so the time average of $(m_z)_{N\to\infty}=Q$
vanishes. 
As a matter of fact, by changing the parameters $A$ and $\omega_0$ of the driving, we would find 
a whole critical line separating regions of phase space where the symmetry is broken from regions
where the symmetry is dynamically restored.
}

It is interesting to explore the connection between thermalization and delocalization of states in the Hilbert space
~\cite{Santos_PRE10,Canovi_PRB11,Marquardt_PRE12,Abanin_AP15}.
An indicator for delocalization is the {\em inverse participation ratio} (IPR)~\cite{Edwards_JPC72} of a
state on a given basis. If we consider the basis $\{|E_n\rangle\}$ of the eigenstates of $\Ham(0)$,
we can study the localization of a given Floquet state $|\phi_{\alpha}(0)\rangle$ by calculating 
$\calI_E(\phi_{\alpha}) =\sum_{n} |\bra{E_n}\left.\phi_\alpha(0)\right\rangle|^4$. 
For the ergodic case, we find a clear delocalization: upon averaging over $\alpha$, we have
$\langle \calI_E(\phi_{\alpha}) \rangle_{\alpha} \sim 1/N$, with fluctuations that scale to $0$ for $N\to \infty$, 
consistently with the Floquet states obeying $T=\infty$-ETH (see~\cite{Russomanno:phdthesis}). 
In the classically regular cases, on the contrary, $\langle \calI_E(\phi_{\alpha}) \rangle_{\alpha}$ is finite and almost
independent of $N$, again with fluctuations scaling to $0$, marking localization of the Floquet states. 
We can gauge delocalization in the Hilbert space~\cite{Anderson_PR58} also by means of
the probability amplitude to remain in $\ket{\psi_0}$, 
$G_{\psi_0}(n\tau) = \bra{\psi_0} \psi(n\tau) \rangle = \bra{\psi_0} \Uevol^n(\tau) \ket{\psi_0}$ 
%
%
whose square $\mathcal{F}_{\psi_0}(t)=|G_{\psi_0}(t)|^2 = |\langle \psi_0 \ket{\psi(t)}|^2$ 
can be seen as a dynamical fidelity \cite{Sharma_EPL14}. 
%
%
Interestingly, one can show that
\begin{equation} \label{F_IPR:eqn}
\overline{\mathcal{F}}_{\psi_0} = \lim_{n\to\infty}\frac{1}{n}\sum_{k=0}^{n-1} | G_{\psi_0}(k\tau) |^2 =
 \sum_{\alpha} \left|R_\alpha\right|^4 \equiv \calI_{F}(\psi_0)\;, \nonumber
\end{equation}
which is the IPR of the initial state $|\psi_0\rangle$ in the Floquet basis.
%
%
In the ergodic cases, we find that $\overline{\mathcal{F}}_{\psi_0}\sim N^{-1}$: 
{\em any} $|\psi_0\rangle$ appears as {\em extended} in the Floquet basis, 
the possibility of Anderson --- here ``dynamical'' ---  localization is excluded, in full agreement with
results on the kicked-top~\cite{Haake_ZPB86}.
Classically regular cases show a diversity of possibilities: some states are definitely localized in the Floquet basis, while
other states show anomalous scaling of the IPR, $\calI_{F}(\psi_0)\sim N^{-\lambda}$ with $0<\lambda<1$, but further
work is necessary to precisely understand the physics behind this.

{\it Conclusions 
{and perspectives}.} We have shown how, in a periodically driven fully connected spin model, 
classical ergodicity translates, at the quantum level, into the system heating up to $T=\infty$,
with Floquet states obeying $T=\infty$-ETH and being delocalized in the Hilbert space.  
On the contrary, if the classical dynamics is regular no thermalization occurs.
We expect that a similar behaviour can be seen in other fully-connected models like 
Bose-Hubbard or Dicke models~\cite{Sciolla_JSTAT11}.
We can propose an experimental set-up to verify our predictions: modulating the inter-well 
barrier height in a double-well BEC realization~\cite{Albiez_PRL05} a driven two-mode Bose Hubbard Hamiltonian
equivalent to our model~\cite{Khripkov_PRE13} can be realized.
Due to ergodicity and the exponential separation of trajectories, 
quantum effects can be easily observed, becoming evident after a few periods even if $N$ is 
large~\cite{Russomanno:phdthesis} and we are well inside the semi-classical regime.
{In this work we have considered the signatures of quantum many body ergodicity manifesting in the behaviour of
{\em local} observables, which have a clear classical limit. The next step is to study the behaviour of genuinely quantum {\em non-local}
objects like the entanglement entropy whose analysis in connection to quantum phase transitions in 
static fully-connected spin chains has been carried out in Refs.~\cite{Filippone_PRA11,Latorre_PRL06}. } 
%
%

We acknowledge useful discussions with G. Biroli, E.G. Dalla Torre, M. Fabrizio, A. Polkovnikov, A. Silva and V. Smelyanskiy.
Research was supported by the Coleman-Soref foundation, by MIUR, 
through PRIN-20087NX9Y7 and RPIN-2010LLKJBX, by SNSF, 
through SINERGIA Project CRSII2 136287\ 1,
by the EU-Japan Project LEMSUPER, and by the EU FP7 under grant 
agreements n. 280555,  n. 600645 (IP-SIQS), n. 618074 (STREP-TERMIQ) .


\newpage
%
%
\section{Supplementary Material: Scaling of the fluctuations}
\begin{figure}
\begin{center}
   \begin{tabular}{c}
    \resizebox{80mm}{!}{\includegraphics{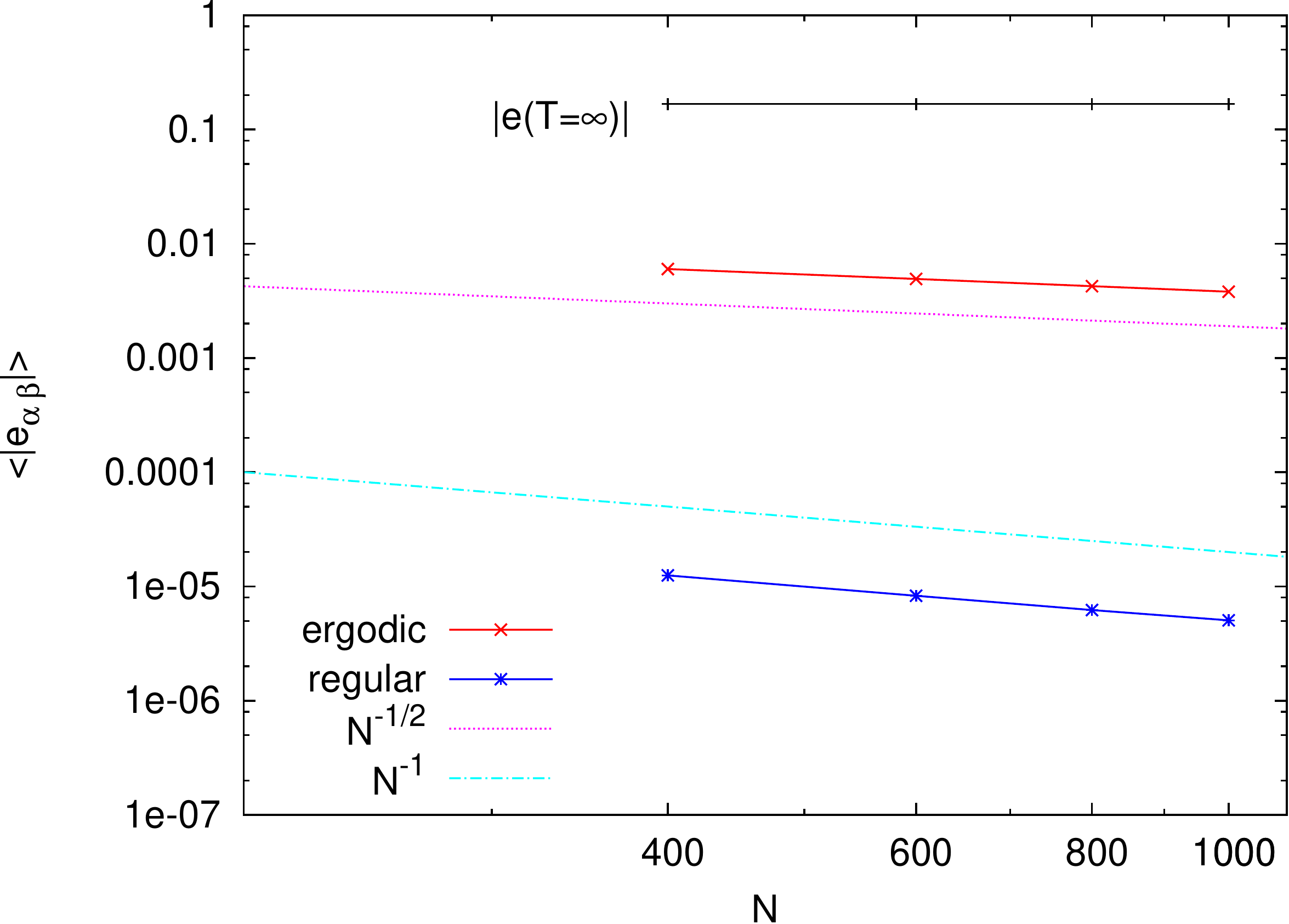}}\\
    \resizebox{80mm}{!}{\includegraphics{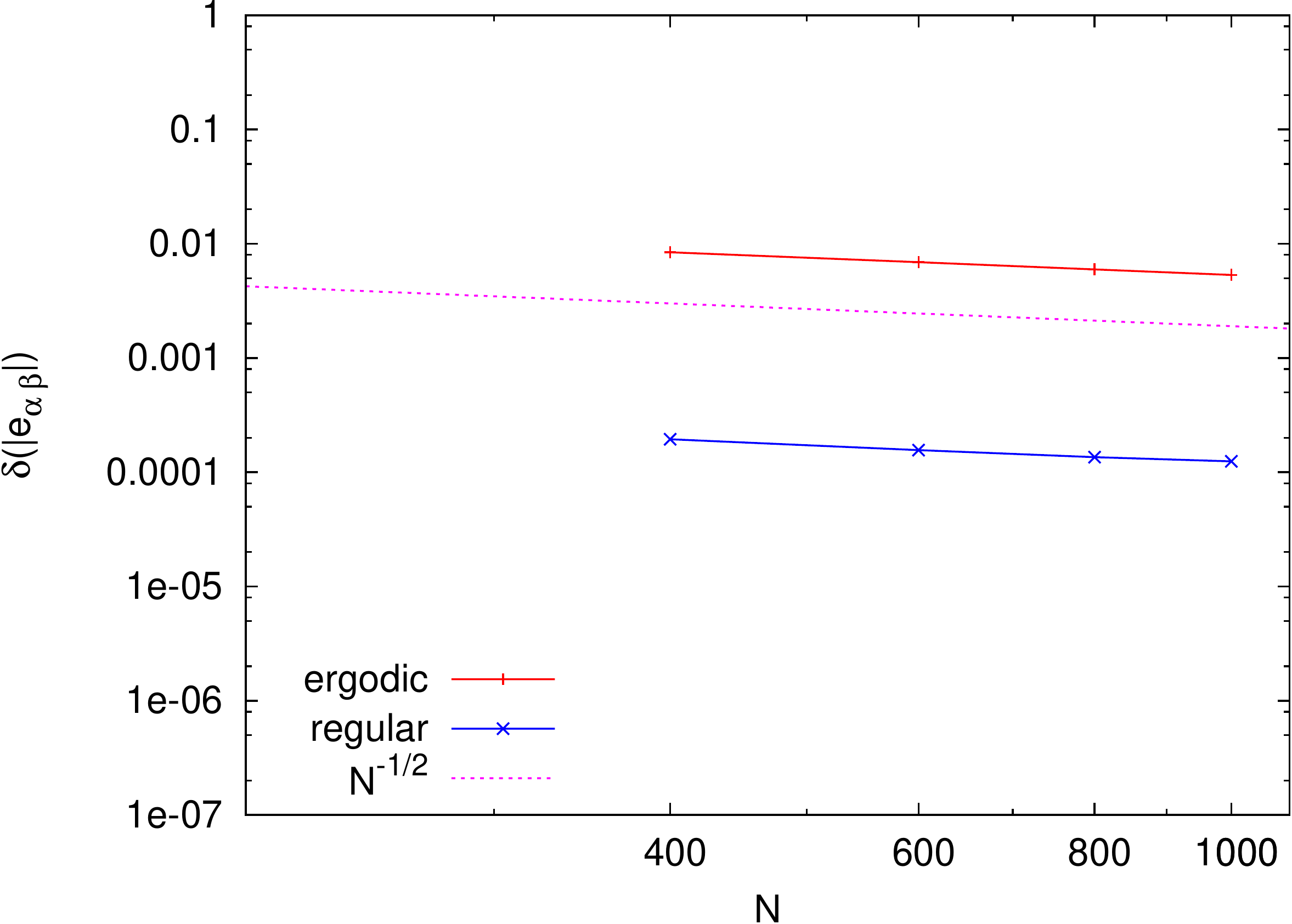}}\\
   \end{tabular}
\end{center}
\caption{(Upper panel) Average of the $|e_{\alpha\beta}|$ vs $N$ in the regular and in the ergodic case. 
         (Lower panel) Fluctuations of the $|e_{\alpha\beta}|$ in the regular and the ergodic case.}
\label{five:fig}
\end{figure}

%
We show in the upper panel Fig.~\ref{five:fig} the scaling of the average off-diagonal terms giving rise to the fluctuations 
(Eq.~(6) of the main text)
$$
  \overline{|e_{\alpha\beta}|} = \frac{1}{N(N+1)}\sum_{\alpha\neq\beta}|e_{\alpha\beta}|\,,
$$ 
both in the regular and the ergodic case. On the lower panel we show the
scaling of the fluctuations of the off-diagonal terms
$$
\delta(|e_{\alpha\beta}|)=\sqrt{\overline{|e_{\alpha\beta}|^2}-\overline{|e_{\alpha\beta}|}^2}\,.
$$
We see that in the regular case these fluctuations are larger: for that reason, in this case,
$\delta e_{\psi_0}$ scales to zero in a less regular fashion with a scaling exponent dependent on
$\ket{\psi_0}$,
as we can see in the bottom-right panel of Fig.~3 of the main text.

When the dynamics is ergodic, we can show analytically that $\delta e_{\psi_0}\sim 1/\sqrt{N}$
whatever is the initial state. The argument runs as follows.
Because in the ergodic case the Floquet states obey ETH at $T=\infty$ (Eq.~(7) of the main text), they can be written as uniform superpositions with random phases of the eigenstates of the Hamiltonian
\begin{equation} \label{randomness:eqn}
  \ket{\phi_\alpha} = \frac{1}{\sqrt{N+1}}\sum_{n=1}^{N+1}\nep^{-i\theta_n^\alpha}\ket{n}
\end{equation}
where $\theta_n^\alpha$ are independent random variables distributed uniformly in $[0,2\pi]$. 
Exploiting that $\ket{n}$ are eigenstates of $\hat{H}(0)$, the Floquet off-diagonal terms $e_{\alpha\beta}(0)$ 
entering in Eq.~(6) of the manuscript can be written as
\begin{equation}
  e_{\alpha\beta}(0) = \frac{1}{N+1}\sum_n \nep^{i\left(\theta_n^\alpha-\theta_n^\beta\right)}e_{nn} \,.
\end{equation}
where $e_{nn}=\langle n|\hat{H}(0)|n\rangle/N$ do not scale with $N$. 
Because the $e_{nn}$ are real, we can separate real and imaginary part of $e_{\alpha\beta}(0)$
\begin{eqnarray}
    \Real e_{\alpha\beta}(0)&=&\frac{1}{N+1}\sum_n \cos{\left(\theta_n^\alpha-\theta_n^\beta\right)}\, e_{nn}\nonumber\\
    \Aimag e_{\alpha\beta}(0)&=&\frac{1}{N+1}\sum_n \sin{\left(\theta_n^\alpha-\theta_n^\beta\right)}\, e_{nn}\,.
\end{eqnarray}
Since $\theta_n^\alpha$ and $\theta_n^\beta$ are distributed uniformly in $[0,2\pi]$, then the distributions of $\cos{\left(\theta_n^\alpha-\theta_n^\beta\right)}$ and $\sin{\left(\theta_n^\alpha-\theta_n^\beta\right)}$ 
are symmetric around zero with a certain {\em finite} variance smaller than $1$. 
By applying the central limit theorem, we see that both the real and the imaginary part of
$e_{\alpha\beta}(0)$ are random variables with vanishing average and variance
scaling like $\sim 1/N$. 
(The situation does not change by taking into account that $e_{nn}$ can have different signs.) 
From this fact it easily follows that $\mean{|e_{\alpha\beta}|^2}\sim1/N$,
where the average is performed on the random distribution of the $\theta_n^\alpha$. 
Considering the overlaps $R_\alpha$, whose square also appears in the fluctuations, Eq.~(6) of the main text, we find
\begin{equation} \label{ralph:eqn}
  R_\alpha = \left\langle\psi_0\right.\ket{\phi_\alpha(0)} = \frac{1}{\sqrt{N+1}}\sum_n \psi_n^*\nep^{-i\theta_n^\alpha}\,.
\end{equation}
where the initial state is expanded in the energy basis as $\ket{\psi_0}=\sum_n\psi_n\ket{n}$.
We have to distinguish two cases: {\em i)} when the initial state is localized in the energy basis and {\em ii)}
when it is extended.
\begin{itemize}
\item[{\em i)}] In the first case, the number of states $|n\rangle$ such that $\psi_n\neq 0$ 
does not scale with $N$: we can schematize this situation by assuming 
$\psi_{\overline{n}}=1$ for some $\overline{n}$ and $\psi_n=0$ for $n\neq\overline{n}$. 
So $|R_\alpha|^2=|R_\beta|^2\approx\frac{1}{N+1}$ and the fluctuations (Eq.~(6) of the manuscript) are
\begin{eqnarray} \label{scal1:eqn}
  &&\hspace{-0.7cm}\delta e_{\psi_0}^2\simeq\frac{1}{(N+1)^2}\sum_{\alpha\neq\beta}|e_{\alpha\beta}|^2\\
  &&\hspace{-0.7cm}\approx 
     \frac{1}{(N+1)^2}\sum_{\alpha\neq\beta}\mean{|e_{\alpha\beta}|^2}\sim \frac{N(N-1)}{N(N+1)^2}\sim\frac{1}{N}\,.\nonumber
\end{eqnarray}
In the second equality we have exploited that the Floquet states are random and self-averaging to replace 
$|e_{\alpha\beta}|^2$ with its average over the probability distribution, $\mean{|e_{\alpha\beta}|^2}$,
whose scaling like $\sim 1/N$ we exploit in the third equality. In the fourth equality we have used that
the number of terms in the sum is $N(N-1)$. 
%
\item[{\em ii)}] Because we are only interested in scalings, we can schematize this situation by assuming that, for all $n$, 
$\psi_n\approx \frac{1}{\sqrt{N+1}}$. 
By applying the central limit theorem to the $R_\alpha$ in Eq.~\eqref{ralph:eqn}, we see
that they are random variables with vanishing average and variance scaling like $1/N$. Hence, the fluctations in Eq.~(6)
of the manuscript would be given by
\begin{eqnarray} \label{scal2:eqn}
  &&\hspace{-0.7cm}\delta e_{\psi_0}^2=\sum_{\alpha\neq\beta}|R_\alpha|^2|R_\beta|^2|e_{\alpha\beta}|^2\\
  &&\hspace{-0.7cm}\leq
    \sum_{\alpha\neq\beta}\mean{|R_\alpha|^2}\mean{|R_\beta|^2}\mean{|e_{\alpha\beta}|^2}\sim \frac{N(N-1)}{N^3}\sim\frac{1}{N}\,,\nonumber
\end{eqnarray}
where the inequality comes from the fact that we are neglecting correlations between the $R_\alpha$ and $e_{\alpha\beta}$. 
%
\end{itemize}
\end{document}